\documentclass[a4paper]{article}
\usepackage{amssymb}

\input{tcilatex}
\begin{document}

\title{Comment on \ \textquotedblleft Bell inequalities and quantum
mechanics\textquotedblright\ by J. H. Eberly}
\author{Luiz Carlos Ryff \\
\textit{Instituto de F\'{\i}sica, Universidade Federal do Rio de Janeiro, }\\
\textit{Caixa Postal 68528, 21941-972 RJ, Brazil}\\
E-mail{\small : ryff@if.ufrj.br}}
\maketitle

\begin{abstract}
Errors in Eberly's derivation of several Bell inequalities are pointed out:
(1) it is based on an equation that is incorrect; (2) it uses neither
two-particle states nor locality to derive Bell's inequalities and; (3) it
does not use entanglement to obtain violations of Bell's inequalities. Even
leading and outstanding\textrm{\ }physicists -- as certainly is the case of
Prof. Eberly -- sometimes make elementary mistakes, and this by no means
diminishes the importance of their scientific contribution. In general, this
is a consequence of an excessive attachment to an idea (nowadays it has
become fashionable to be against realism). This shows the importance of
trying to keep an open mind.
\end{abstract}

In an article on Bell inequalities and quantum mechanics \textrm{[1]},
Eberly uses an arrangement of optical loops to derive several Bell
inequalities. As we will see, his reasoning is incorrect, and his
conclusion, according to which \emph{\textquotedblleft there really isn't a
sound when a tree falls if there is no way to record it,\textquotedblright }%
\ is unfounded \textrm{[2]}.

Eberly's set-up uses \emph{\textquotedblleft five analyzer loops, two to the
left and three to the right of a photon source.\textquotedblright }\ Each
loop is constituted by \emph{\textquotedblleft a pair of birefringent
crystals arranged with an air gap between them, and cut and positioned in
such a way that a light beam entering the first crystal is divided into
orthogonally polarized components that travel separately across the air gap
and are then recombined into the original beam by the second
crystal.\textquotedblright\ }The source emits polarization entangled photons
in the state $(1/\sqrt{2})(\mid x\rangle \mid y\rangle -\mid y\rangle \mid
x\rangle )=(1/\sqrt{2})(\mid \theta \rangle \mid \overline{\theta }\rangle $

\noindent $-\mid \overline{\theta }\rangle \mid \theta \rangle )=(1/\sqrt{2}%
)(\mid \phi \rangle \mid \overline{\phi }\rangle -\mid \overline{\phi }%
\rangle \mid \phi \rangle $ \textrm{[3]}, where the bar \emph{%
\textquotedblleft is used to denote orthogonal complement. For example, }$%
\overline{\theta }=\theta \pm 90^{\circ }$\emph{, and all angles are
measured from the x axis.\textquotedblright\ }The experiment takes place in
three stages. In stage 1 \emph{\textquotedblleft The experimenter records
the fraction of times a photon is detected on the right, given the detection
of a y-polarized photon on the left. This fraction will be designated as }$%
f(x,\phi )$\emph{\ to indicate that the right-moving photon was originally
x-polarized but was detected as }$\phi $\emph{-polarized (necessarily so,
because the }$\overline{\phi }$\emph{\ channel was
blocked).\textquotedblright }\ I will designate this fraction as $%
f_{1}(x,\phi )$, to make it clear that it is recorded in stage 1. In stage 2 
\emph{\textquotedblleft The experimenter records the fraction of times a
photon is detected on the right, given the detection of the x-polarized
photon on the left. This fraction will be designated as }$f(y,\theta )$\emph{%
\ to indicate that the right-moving photon was originally y-polarized but
was detected as }$\theta $\emph{-polarized (necessarily so, because the }$%
\overline{\theta }$\emph{\ channel was blocked).\textquotedblright } I will
designate this fraction as $f_{2}(y,\theta )$, to make it clear that it is
recorded in stage 2. In stage 3 \emph{\textquotedblleft The experimenter
records the fraction of times a photon is detected on the right, given the
detection of the }$\overline{\theta }$\emph{-polarized photon on the left.
This fraction will be designated as }$f(\theta ,\phi )$\emph{\ to indicate
that the right-moving photon was originally }$\theta $\emph{-polarized but
was detected as }$\phi $\emph{-polarized (necessarily so, because the }$%
\overline{\phi }$\emph{\ channel was blocked).\textquotedblright } I will
designate this fraction as $f_{3}(\theta ,\phi )$, to make it clear that it
is recorded in stage 3.

According to Eberly, in stage 1 it \textit{appears} to be obviously true
that \emph{\textquotedblleft Because we do not ask which of the }$\theta $ 
\emph{or }$\overline{\theta }$ \emph{channels any of those photons went
through in traversing the intermediate loop, we }[can] \emph{decompose }$%
f(x,\phi )$ \emph{to include both possibilities, which we indicate by
writing }$f(x,\phi )=f(x,\theta ,\phi )+f(x,\overline{\theta },\phi )$\emph{%
.\textquotedblright } But Eberly thinks this is actually a mistake that
leads to the puzzling features of the first Bell inequality he derives.
However, his reasoning is incorrect. To verify this, the above expression
can be rewritten as 
\begin{equation}
f_{1}(x,\phi )=f_{1}(x,\theta ,\phi )+f_{1}(x,\overline{\theta },\phi ). 
\tag{1}
\end{equation}%
Applying the same \textquotedblleft mistaken\textquotedblright\ reasoning to
stage 2, Eberly decomposes $f(y,\theta )$ by writing $f(y,\theta
)=f(y,\theta ,\phi )+f(y,\theta ,\overline{\phi })$. This can be rewritten
as 
\begin{equation}
f_{2}(y,\theta )=f_{2}(y,\theta ,\phi )+f_{2}(y,\theta ,\overline{\phi }). 
\tag{2}
\end{equation}%
Applying the \textquotedblleft mistaken\textquotedblright\ reasoning to
stage 3, Eberly decomposes $f(\theta ,\phi )$ by writing $f(\theta ,\phi
)=f(x,\theta ,\phi )+f(y,\theta ,\phi )$, which can be rewritten as%
\begin{equation}
f_{3}(\theta ,\phi )=f_{3}(x,\theta ,\phi )+f_{3}(y,\theta ,\phi ).  \tag{3}
\end{equation}

Following his line of reasoning, Eberly writes: \emph{\textquotedblleft By
direct addition of the numbers of photons in the categories described, we
see that }%
\begin{equation}
f(x,\phi )+f(y,\theta )=f(x,\theta ,\phi )+f(x,\overline{\theta },\phi
)+f(y,\theta ,\phi )+f(y,\theta ,\overline{\phi }).  \tag{4}
\end{equation}%
\emph{It is simple to observe that among the terms on the right-hand side of
Eq. (4) }[Eq. (1) in the original text]\emph{, we find both }$f(x,\theta
,\phi )$ \emph{and }$f(y,\theta ,\phi )$\emph{,} \emph{and the sum of them
is }$f(\theta ,\phi )$\emph{. That is, another way to write Eq. (4) is }%
\begin{equation}
f(x,\phi )+f(y,\theta )=f(\theta ,\phi )+f(x,\overline{\theta },\phi
)+f(y,\theta ,\overline{\phi }).  \tag{5}
\end{equation}%
\emph{If we drop the two final terms (both are positive or zero fractions),
we obtain the following inequality: }%
\begin{equation}
f(x,\phi )+f(y,\theta )\geqslant f(\theta ,\phi ),  \tag{6}
\end{equation}%
\emph{which is an example of what is called a Bell inequality, after the
physicist John Bell, who first studied their consequences in quantum physics
in the mid-1960s.\textquotedblright }

And now we can see where Eberly's mistake lies. Eq. $(4)$ can be rewritten
as 
\begin{equation}
f_{1}(x,\phi )+f_{2}(y,\theta )=f_{1}(x,\theta ,\phi )+f_{1}(x,\overline{%
\theta },\phi )+f_{2}(y,\theta ,\phi )+f_{2}(y,\theta ,\overline{\phi }), 
\tag{7}
\end{equation}%
and there is no justification to assume that $f_{1}(x,\theta ,\phi
)+f_{2}(y,\theta ,\phi )=f_{3}(x,\theta ,\phi )+f_{3}(y,\theta ,\phi
)=f_{3}(\theta ,\phi )$, as Eberly did, where $f_{1}(x,\theta ,\phi )$
correspond to right-moving photons that were originally x-polarized, $%
f_{2}(y,\theta ,\phi )$ to right-moving photons that were originally
y-polarized, and $f_{3}(x,\theta ,\phi )$ and $f_{3}(y,\theta ,\phi )$ to
right-moving photons that were originally $\theta $-polarized. \emph{%
\textquotedblleft A variety of inequalities that are similar to Eq. (6) }%
[Eq. (3) in the original text]\emph{\textquotedblright\ }derived by Eberly
are based on the same incorrect reasoning.

It is also interesting to observe that Eberly's argument actually involves
only single particle states. We can simply ignore the two analyzer loops to
the left and send right-moving x-polarized photons in stage 1, right-moving
y-polarized photons in stage 2, and right-moving $\theta $-polarized photons
in stage 3. This makes it clear that no authentic Bell inequality can be
obtained in this way, since Bell's argument requires two-particle states and
the locality assumption. Moreover, although we don't need entanglement to
derive Bell's inequalities -- as correctly emphasized by Eberly --
entanglement is an essential ingredient in obtaining violations of Bell
inequalities.{\huge \ }

I would like to add that, in principle, Eberly's experiment is also valid
for material particles with spin \textrm{[4] }(we only have to use
Stern-Gerlach apparatuses to split the beams and magnetic fields to
recombine them), and that it can be explained by Bohmian mechanics; with no
need, however, it is important to stress, of taking into consideration
Bohmian mechanics' nonlocal features \textrm{[5]}.

Just for the sake of completeness, I will introduce a realistic model, based
on the pilot wave interpretation \textrm{[5,6]}, that explicitly
demonstrates the incorrectness of Eberly's standpoint.\textrm{\ }In $(1)$ we
have $f_{1}(x,\theta ,\phi )=\cos ^{2}\theta \cos ^{2}\phi $, where $\cos
^{2}\theta $ is the probability of having the photon follow the $\theta $
channel and an empty wave follow the $\overline{\theta }$ channel, and $\cos
^{2}\phi $ is the probability of having the photon follow the $\phi $
channel. Similarly, $f_{1}(x,\overline{\theta },\phi )=\sin ^{2}\theta \cos
^{2}\phi $, where $\sin ^{2}\theta $ is the probability of having the photon
follow the $\overline{\theta }$ channel. Applying the same reasoning, in $%
(2) $ we have $f_{2}(y,\theta ,\phi )=\sin ^{2}\theta \cos ^{2}(\phi -\theta
)$ and $f_{2}(y,\theta ,\overline{\phi })=\sin ^{2}\theta \sin ^{2}(\phi
-\theta )$, and in $(3)$ we have $f_{3}(x,\theta ,\phi )=\cos ^{2}\theta
\cos ^{2}(\phi -\theta )$ and $f_{3}(y,\theta ,\phi )=\sin ^{2}\theta \cos
^{2}(\phi -\theta )$ \textrm{[7]}.


\begin{thebibliography}{9}
\bibitem{2} The complete passage reads: \textquotedblleft But apart from all
its striking calculational successes, quantum theory brings more to physics,
namely a world view \ that can be quite uncomfortable. To adopt an extreme
phrasing, there really isn't a sound when a tree falls if there is no way to
record it.\textquotedblright
\bibitem{7} Please, note that these are general conditions that a local
model should fulfil, assuming ideal conditions (e.g., ideal detectors, and
so on). They make it clear that -- contrary to his claims -- Eberly's
experiment can be interpreted from a local realistic standpoint.
\end{thebibliography}
\end{document}